\documentclass[runningheads,a4paper]{llncs}

\usepackage{amssymb}
\usepackage{color}

\usepackage{graphicx}
\usepackage{float}
\usepackage{pdfcomment}
\usepackage[capitalise]{cleveref}
\usepackage{cite}

\usepackage{url}
\urldef{\mailsa}\path|{kfeldman,lfaust,xwu9,chuang7,nchawla}@nd.edu|
\newcommand{\keywords}[1]{\par\addvspace\baselineskip
\noindent\keywordname\enspace\ignorespaces#1}

\begin{document}

\mainmatter  % start of an individual contribution

% first the title is needed
\title{Beyond Volume: \\ The Impact of Complex Healthcare Data \\ on the Machine Learning Pipeline}

% a short form should be given in case it is too long for the running head
\titlerunning{The Impact of Complex Healthcare Data on the Machine Learning Pipeline}

% the name(s) of the author(s) follow(s) next
%
% NB: Chinese authors should write their first names(s) in front of
% their surnames. This ensures that the names appear correctly in
% the running heads and the author index.
%
\author{Keith Feldman\inst{1}
\ Louis Faust\inst{1}\ Xian Wu\inst{1}\\ Chao Huang\inst{1} and Nitesh V. Chawla \inst{1}}
\authorrunning{Feldman et al.}
% (feature used for this document to repeat the title also on left hand pages)

% the affiliations are given next; don't give your e-mail address
% unless you accept that it will be published
\institute{University of Notre Dame, Notre Dame, IN, USA\\
\mailsa\smallskip
%\and
%AIT Austrian Institute of Technology GmbH, Health \& Environment Department, %Biomedical Systems, Donau-City-Str. 1, A-1220 Vienna, Austria\\
%\mailsb\smallskip
}

%
% NB: a more complex sample for affiliations and the mapping to the
% corresponding authors can be found in the file "llncs.dem"
% (search for the string "\mainmatter" where a contribution starts).
% "llncs.dem" accompanies the document class "llncs.cls".
%

\toctitle{Beyond Volume: The Impact of Complex Healthcare Data on the Machine Learning Pipeline}
\tocauthor{Feldman et al.}
\maketitle

\begin{abstract}
From medical charts to national census, healthcare has traditionally operated under a paper-based paradigm. However, the past decade has marked a long and arduous transformation bringing healthcare into the digital age. Ranging from electronic health records, to digitized imaging and laboratory reports, to public health datasets, today, healthcare now generates an incredible amount of digital information. Such a wealth of data presents an exciting opportunity for integrated machine learning solutions to address problems across multiple facets of healthcare practice and administration. Unfortunately, the ability to derive accurate and informative insights requires more than the ability to execute machine learning models. Rather, a deeper understanding of the data on which the models are run is imperative for their success. While a significant effort has been undertaken to develop models able to process the volume of data obtained during the analysis of millions of digitalized patient records, it is important to remember that volume represents only one aspect of the data. In fact, drawing on data from an increasingly diverse set of sources, healthcare data presents an incredibly complex set of attributes that must be accounted for throughout the machine learning pipeline. This chapter focuses on highlighting such challenges, and is broken down into three distinct components, each representing a phase of the pipeline. We begin with attributes of the data accounted for during preprocessing, then move to considerations during model building, and end with challenges to the interpretation of model output. For each component, we present a discussion around data as it relates to the healthcare domain and offer insight into the challenges each may impose on the efficiency of machine learning techniques.
  

\end{abstract}

\section{Introduction}
Only in its infancy as a digital entity, the healthcare industry has undergone a significant transition over the past decade from a paper-based domain to one operating primarily through a digital medium. Beyond the logistical benefits of maintaining and organizing patients' medical records, the ability to quickly identify and process information from millions of patient records, laboratory reports, imaging procedures, payment claims, and public health databases has brought the industry to the precipice of a significant change. Namely, the opportunity to utilize data science and machine learning methodologies to address problems across the practice and administration of healthcare.

In fact, utilization of such analytic techniques has provided a foundation on which models of personalized and predictive care have emerged\cite{yoo2012data}. These models represent a myriad of opportunities from improved patient stratification, to identifying novel disease comorbidities and drug interactions, to the prediction of clinical outcomes~\cite{jensen2012mining}. However, while such applications hold great promise for the healthcare industry, the application of machine learning methodologies faces a significant set of obstacles intrinsic to the data being evaluated and the population from which the data is drawn.

Since its entrance into the digital era, the increasing scale and scope of data has placed great emphasis on the advent of \textit{Big Data} in healthcare and the challenges that come with it. With an estimated 150 exabytes of data generated by 2011, early work addressed the challenges of processing data at such a scale~\cite{hughes2011big}. However, it is important to remember that Big Data is defined by more than just size, but rather by what are known as the four V's (The \textbf{V}olume, or quantity of data available. The \textbf{V}elocity, or speed at which the data is created. The \textbf{V}ariety of the data elements available. And the \textbf{V}eracity, or inherent truthfulness of data itself)~\cite{raghupathi2014big}. With advancements to the theoretical underpinning and practical implementations of machine learning algorithms providing the ability to consume and analyze even the largest clinical and biomedical datasets, the challenge now falls not to the size of the data, but its complexity. 

In stark contrast to the idealistic data on which machine learning algorithms are theoried, healthcare data is inherently fragmented, noisy, high-dimensional, and heterogeneous. With the influx of data from an increasingly varied set of sources, it has become clear that effective utilization of these techniques will require more than accessibility of data or ability to execute Big Data analytics. As clinical research becomes increasingly intertwined with the statistical methodologies of data science, effective applications require an awareness to the mechanisms by which the data is created, processed, and analyzed. 

\begin{figure}
\small
\centering
\includegraphics[width=1\textwidth]{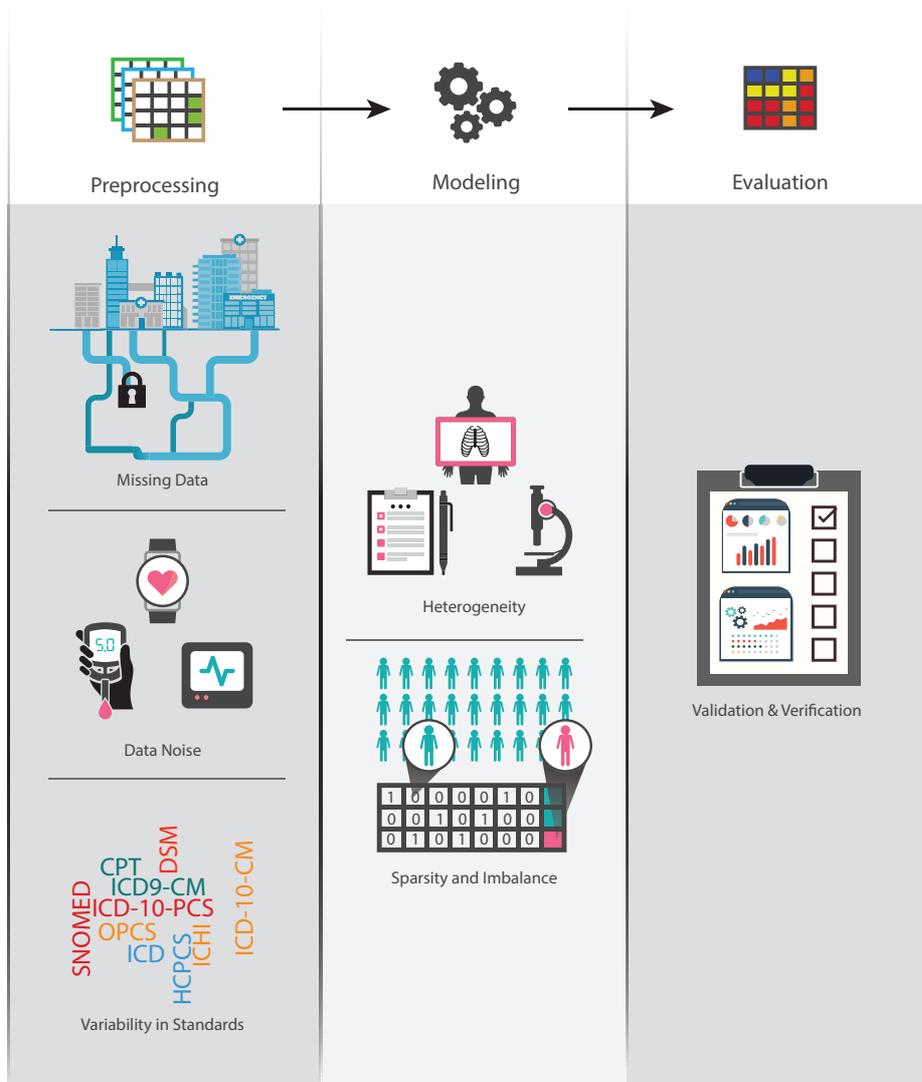}
 \caption{Fundamental challenges to the machine learning pipeline resulting from the complexity of healthcare data}
 \label{figure:pipeline}
\end{figure}

To this end, the following chapter will address the complexities of healthcare data as they impact the machine learning algorithms which consume them. Broadly, we break such work into three major categories, each representing a component of the machine learning pipeline, as seen in Figure~\ref{figure:pipeline}. Beginning with preprocessing, we will discuss attributes of the data itself through the concepts of noise, missingness, and variability in language. We will then move to the modeling phase, discussing considerations such as the heterogeneity of data sources, sparsity and class imbalance. Finally, we will look to the model output, discussing the concepts of validation and verification. We will conclude with some general recommendations and a review of the open problems. 

\section{Glossary and Key Terms}

\noindent\textit{\underline{Preprocessing}}: A process intended to address the noisy, missing, and inconsistent properties of real-world data, improving data quality prior to modeling. Preprocessing is often the first step in the machine learning pipeline and is characterized by techniques such as cleaning, integration, reduction, and transformation~\cite{han2011data}. \medskip

\noindent\textit{\underline{Modeling}}: The second stage of the machine learning pipeline, modeling focuses on the construction of statistical, probabilistic models intended to learn representations of the vast amounts of data collected. Such models are used to detect patterns in data and potentially use the patterns to predict future data~\cite{murphy2012machine}. \medskip

\noindent\textit{\underline{Evaluation}}: Performed on the artifacts produced by modeling, evaluation forms the final stage of the pipeline, establishing the model's predictive efficacy, complexity, technical correctness, and ease with which it can be understood~\cite{sammut2011encyclopedia}. \medskip

\noindent\textit{\underline{Validation}}: The process of evaluating a model in its ability to accurately represent the observed system~\cite{kantardzic2011data}. \medskip

\noindent\textit{\underline{Verification}}: The process of evaluating whether data manipulation and model construction were accomplished with technical correctness~\cite{kantardzic2011data}.\medskip

\noindent\textit{\underline{Medical Coding}}: The systematic classification of data into alphanumeric codes for the identification of diagnoses, procedures, medications, laboratory tests, and other clinical attributes~\cite{diamond2013mastering}.\medskip

\noindent\textit{\underline{Sparsity}}: Occurs when only a small percentage (typically $<1$\%) of attributes for an instance are non-zero~\cite{tan2006introduction}. \medskip

\noindent\textit{\underline{Concept Drift}}: The notion that inputs or outcomes related to a model may change overtime in unexpected manners reducing the accuracy of models as the data streams change~\cite{tsymbal2004problem}. 

\section{Preprocessing}

Just as clinicians require quick and accurate information to provide care at the highest level, the need to collect and produce high quality data has become paramount for applications of machine learning as they continue to integrate into aspects of care pertaining to health and human behavior. However, while the goal is clear, the rapid influx of new data, and the evolving nature of healthcare itself offers a significant set of challenges. In this regard, the following sections present an overview to the considerations of preprocessing data collected across the healthcare domain. 

\subsection{Manifestation in Healthcare}
While the challenges to preprocessing are present in many domains, the dynamics of healthcare necessitate that care be taken to address a number of biological, computational, and representational aspects of data. These can range from the filtration of noise, to the need to navigate a multitude of coding standards. The following sections will begin by highlighting scenarios from which these challenges arise.

\subsubsection{Noise}

The presence of incorrect or irrelevant data, otherwise known as noise, represents a fundamental component of working with any real-world data. Healthcare is no exception, and arising from an imperfect data collection process, common occurrences of noise can include missing values, misspellings, abbreviations, misfielded values, word transpositions, and duplicated or conflicting records \cite{rahm2000data}. The presence of noise stems from more than just data quality issues, it may also arise from the natural variation among individuals. Given a population of instances, a small sample may appear inconsistent with the rest i.e. ``outliers''.

Next, it is important to remember that noise is present not only in the recording of data, but in its measurement as well. Healthcare is currently entering uncharted territory. While traditionally the generation of health data was confined within the walls of a clinical setting, advancing technology has allowed for collection from a variety of sources. These range in complexity from personal health tools, to clinically focused devices, to total wireless sensor networks, to home monitoring systems~\cite{king2002digital, andreu2015wearable, kidd1999aware}. However, with development from a number of manufacturers, utilizing a range of algorithmic techniques for their data collection and approximation, the quality of this data has been drawn into question \cite{caceres1983medical, koumoundouros2014clinical}. A scenario highlighted by Bland and Altman, who note that ``several measurements of the same quantity on the same subject will not in general be the same. This may be because of natural variation in the subject, variation in the measurement process, or both'' \cite{bland1996statistics}.  

Finally, we find that beyond the collection and recording of health data, there exists a more complex source of noise known as artifacts. Artifacts result not from data collection or variability in subjects, but from the physiological processes which generate the data itself, manifesting as what appears to be normal data. Although, in reality, such feature values are not generated by the intended source (e.g. electrical signals from the brain collected by an electroencephalogram). Instead, this data is generated by alternative biological mechanisms including cardiac, glossokinetic, muscle, eye movements, respiratory and pulse variations~\cite{sethi2007physiological}.

\subsubsection{Missingness}
The occurrence of missing data is an almost unavoidable problem for any domain, including healthcare~\cite{wood2004missing}. Missing data can result from a number of processes, ranging from fundamental attributes of data collection to the inherent ambiguity and variability of an individual's health condition. At the most basic level, as with all studies that involve the collection of information from individuals, there exists the possibility of missing data attributes due to a subjects failure to respond completely, as well as the inability to assess all possible clinical and social attributes as they pertain to each individual. Additionally, missing data is not restricted to particular attributes, but can arise on a broader scale with attrition of an entire instance during longitudinal data collection. In addition to the to the lack of data, such a scenario presents difficultly during processing as the reason for dropping may be linked to attributes of the study design, a trend which may go unnoticed without closer investigation~\cite{little2012prevention}. The evolving digital nature of healthcare presents its own set of challenges in regards to the presence of missing data. From a collection standpoint, monitoring devices may fail or become disconnected, data may become corrupt, or compatibility issues may result in the inability to collect data, resulting in large gaps of the recorded data~\cite{marlin2012unsupervised}. 

Moreover, even in the scenario in which data collection occurs as expected, missingess can take other forms. Looking to the frequency of clinical encounters, it may occur from a temporal standpoint. With the exception of some critical care, patients are rarely under continuous observation, instead, many may meet with their physicians as infrequently as twice a year. These gaps in observation and records may allow for fluctuations in health to go undocumented, leaving only brief snapshots of the patients condition. 

Further, missingness can occur due to the fragmented nature of the entities collecting the data.  Healthcare data comes not only from hospitals and primary care centers, but a variety of sources, be that specialists visited, community programs, or even physical trackers~\cite{azarm2015review, andreu2015wearable}. However, despite the various data sources collecting data relevant to the overall profile of an individuals health, data integration and sharing considerations often provide only a small portion of data to any one source. 
  
Finally, it is important to note a distinction between missing data and negative values as it relates to a perhaps non-traditional concept of missingness. Unlike domains such as retail, where the purchase of an item can be represented in a binary fashion (purchased or not), the lack of affirmation for a particular entity in healthcare data does not necessitate a negative case. Looking to disease diagnoses, a patient may in fact have a particular condition, for which they are never formally diagnosed, or for which the diagnosis code is not recorded, as is the case with often under-reported diagnoses such as obesity~\cite{quan2008assessing}.

\subsubsection{Variability in Language}
Another challenge in processing healthcare data stems not from a function of its quality, but from its representation. In an effort to quantify and standardize the vast set of possible conditions, procedures, and clinical elements, a myriad of medical coding schemes have been developed, including the International Classification of Diseases (ICD), Systematized Nomenclature of Medicine (SNOMED), Current Procedural Terminology (CPT), Healthcare Common Procedure Coding System (HCPCS), LOINC, Europe's Classification of Surgical Operations and Procedures (OPCS), and the Diagnostic and Statistical Manual of Mental Disorders (DSM) to name a few. In fact, as the number of standards continues to increase there is a considerable amount of overlap between them. As a result, effective processing of such data must take care to consider the possibility where the same attribute may be represented in multiple ways. This situation is exacerbated by the nature of healthcare systems, where in response to documentation or reporting standards, multiple coding standards may be used even within the same institution. 

Not only does variability arise from the use of different coding standards, but from emerging diversity as these standards are revised and updated. As an example, the ICD's latest revision (ICD-10) brought with it roughly 55,000 new diagnostic codes and over 68,000 new procedural codes~\cite{cdc_icd10}. Although this increased feature space allows for representation of conditions at a much greater specificity, coalescing codes across revisions during processing presents a significant challenge. Further, although mappings have been created to assist in the transition between codes, they are not univeral and often incomplete. The Workgroup for Electronic Data Exchange suggest ``healthcare organizations use these mappings as starting points to develop their own, more precise data crosswalk applications between ICD-9 and ICD-10 codes''~\cite{meyer2011coding}.

While variability is clearly a product of the expansive set of coding standards and their revisions, it also results from the methodology of medical coding itself. Medical coding is a subjective process, the accuracy of which has been shown to be dependent on the clinical record of the condition observed, as well as the interpretation of the diagnostic codes themselves~\cite{fisher1992accuracy}. While it may be straightforward for simple cases where a patient is assigned a single diagnosis, inconsistencies from coders and institutions have been found to increase with the complexity of a patient's condition, specifically when they receive multiple diagnoses~\cite{macintyre1997accuracy}.

\subsection{Implications to Machine Learning}
Noted by Cortes et. al, ``insufficiencies of the data limit the performance of any learning machine or other statistical tool constructed from and applied to the data collection - no matter how complex the machine or how much data is used to train it'' \cite{cortes1995limits}. As a result, it is imperative to understand not only the processes from which preprocessing challenges arise within the healthcare domain, but also the implications to the preprocessing phase of the machine learning pipeline. A discussion to each consideration can be found in the sections below. 

\subsubsection{Noise}
As applications of machine learning continue to expand into new aspects of healthcare, the processing of noisy data has become a central component of many works. From a theoretical perspective, prior work has established fundamentals of what defines learning, and concepts of model consistency. Together, these constructs help illustrate how failure to process noise in data can cause difficulties in constructing a model that accurately reflects the population from which the data is drawn, negatively impacting generalizable performance~\cite{vapnik1998statistical, kearns1994introduction, sessions2006effects}. 

While much of the standard noise can be attributed to data quality issues, it is important to highlight the need for data understanding in the preprocessing step. In particular, with relation to outliers. There are over 100 different discordancy/outlier tests whose use can depend on factors such as data distribution, whether distribution parameters are known, and even the number and type of the expected outliers~\cite{knorr2000distance}. As such, preprocessing noisy data presents a significant challenge, as incorrectly applying data cleaning techniques can result in large variations in the finalized dataset. 

Looking to other sources of noise, the nature of potential physiological artifacts requires additional considerations during data preprocessing. The presence of an artifact does not necessitate the value be incorrect, though similarly to outliers, failure to remove such data has far reaching implications as such data presents ``a milder form of training data error that can cause reduced accuracy''~\cite{bacioiu2008method}. However, unlike outliers, these artifacts are often difficult to broadly and statistically discern from true signals without clinical insight.

Finally, in addition to the data quality concerns already discussed, noise resulting from the variability of the systems which collect data presents a distinct concern during processing. Due to the resulting intra-instance variability, there exists the case in which two instances with identical feature values present two different classes or outcomes. Whereas such instances are often removed, within healthcare, such a scenario is quite common and may represent a legitimate aspect of variability with a patient's health. This in turn, introduces a considerable amount of uncertainly into the system, representing an inherent problem to separability.  

\subsubsection{Missingness}

Just as the numerous sources of noise present a challenge to the effective processing of healthcare data, so too do the many forms in which missing data can manifest across the domain. At an attribute-level, data is typically classified as missing in one of three forms: completely at random (MCAR), at random (MAR), and not at random (MNAR). Although all forms of missingess present a concern, the various forms of missingness can present significantly different considerations during the processing of a dataset. While data missing completely at random (MCAR) presents minimal concern to the underlying distribution, allowing for data to be dropped or imputed without worry of introducing additional biases, such a scenario is often unrealistic. Rather, data is typically missing due to an underlying, sometimes unobserved, pattern known as missing at random (MAR) or missing not at random (MNAR), each of which may require techniques such as maximum likelihood estimation or multiple imputation to help address the inherent bias they present to the data collected~\cite{little1987statistical,arbuckle1996full, rubin2004multiple}. In comparison, both methods tend to yield similar results when implemented in the same way, however, performance gains in efficiency and reduced bias regarding these methods relate to the inclusion of auxiliary variables~\cite{collins2001comparison}. The inclusion of data to these methods, even that which is irrelevant to the objective at hand, suggests the amount of data included in these methods is of equal importance to the methods themselves.

Beyond the type of missingness, the quantity of missing information further influences the preproccessing of data. With respect to the occurrences of large temporal gaps, we find that although mathematically we may be able to impute, model and predict estimations of missing values during processing there is no guarantee the values computed accurately reflect the true condition of the individual during that time period. This consideration is particularly relevant in light of the common scenario where data is collected during a subjects clinical encounters, each of which may occur months apart.  

Finally, building on the concerns of temporal missingess, commonly associated with longitudinal studies, missingness by attrition, presents a number of additional considerations to effective preprocessing of data. Work by Graham suggests attrition-related missingness focuses on the program (or treatment) \textit{P}, the dependent \textit{Y}, and the interaction between these two: \textit{PY}~\cite{graham2012missing}. Just as the MCAR/MAR/MNAR nomenclature provides a roadmap to the appropriate preprocessing techniques, identifying and assessing which of the possible combinations of these three factors causes missingess to arise presents a critical step in improving the ability to address bias during the processing of such data.

\subsubsection{Variability in Language}
The variability brought on by the breath of coding standards presents a fundamental obstacle in the effective preprocessing of healthcare data. Although an underlying condition may be the same across two distinct representations, with the multitude of values across each of the different coding standards, it has become nearly impossible to accurately create a comprehensive mapping to translate between each standard. However, such a mapping is critical for unifying disparate data sources during the processing stage of the machine learning pipeline.

Further, although data may stay consistent with respect to a single coding standard, temporal changes in how these codes are assigned can still occur as a result of changing regulations, or even revisions within the standard~\cite{rector2008hard}. Presenting similar obstacles as with multiple coding standards, these changes, more formally defined by the notion of concept drift, cause models built on old data to become inconsistent with new data as the models inputs and target variable change over time~\cite{tsymbal2004problem}. 

While such a shift is extremely difficult to identify, it is critically important, as such discrepancies make it not only difficult to understand values, but have the potential to add ambiguity during its processing~\cite{lindenauer2012association}. In particular, changes in the code frequencies, which are often used during preprocessing and data exploration, may be attributed to other clinical attributes, rather than the true shift in language. For example, an individual with a chronic illness may have a record with multiple representations due to changes in how the illness was labeled over time. When such changes go unaccounted for, the record may be perceived as having three distinct illnesses instead of one. 

\section{Modeling}

To this point we have discussed intrinsic characteristics of data, those properties which influence the statistical foundations guiding machine learning theory. However, we now look further, not to the properties of the data, but to the mechanisms through which the data is consumed and represented to build effective machine learning models. Such attributes range from high-level aspects of integrating heterogenous data types, to low-level considerations when representing an increasingly expansive feature space.  

\subsection{Manifestation in Healthcare}
As before, we will begin with an outline of the processes within the healthcare domain from which such considerations arise. An overview of each can be found in the respective sections to follow.

\subsubsection{Heterogeneity}
 
Drawn from multiple sources and encompassing multiple modalities, healthcare data represents a remarkably heterogeneous set of data types and sources. Perhaps the most prominent examples can be found within the wealth of clinical data now digitalized as a result of EMR integration across healthcare practices. Typically, such data is broken into structured data including diagnosis codes, procedural information, medication data, laboratory test results, data recorded directly from patient's bedside monitors, and unstructured data such as images and clinical text~\cite{weber2014finding}. However, data can also include patient demographics, financial claims, and more recently, genomic sequencing and other omics data, each of which may require different considerations as they are processed during modeling.
 
Although electronic health records are perhaps the most well-known source of data, health-related data can be collected, inferred, and analyzed from a number of indirect sources. These can include common population health and reporting fields such as the census bureau and the department of labor statistics, as well as less obvious sources such as the location of fresh food sources in a city. It is also worth noting the number of external data is only expected to increase, as shifts in the regulatory landscape of the healthcare industry have advanced the collection and analysis of population health data though a number of initiatives~\cite{stoto2013population}. 
 
Finally, it is important to note that heterogeneity can exist even within data of the same type. As an example, through prior work our group has established fundamental differences between clinical notes based on the clinical occupation of those who write them~\cite{feldman2016mining}. While important for the processing of clinical text, the establishment of such heterogeneity impresses a deeper need for an awareness of not only the types of data we process, but the varied sources of data from which models are constructed. 

\subsubsection{Dimensionality, Sparsity and Imbalance}
 
Beyond the variety of sources generating data, the digitalization of healthcare data has resulted in a significant increase in the number of features able to be extracted with respect to an instance, i.e. its dimentionality. Although the high-dimensional data resulting from the processing of unstructured images and text has become commonplace, advancements in clinical and computational technology now allow for improved analysis of biological processes at their most basic level. As an example, resulting from the increasing affordability of genomic sequencing, we have witnessed a rise of genome-wide association studies, which aim to represent and identify associations between the over 10 million common single-nucleotide polymorphisms (SNPs) in the human genome~\cite{visscher2012five}. 

However, such expansive feature sets are not only a result of biological mechanisms, but of artificial constructs used to structure the data itself. In particular, we look to the coding standards used to represent attributes of a patient's condition and care. As noted prior, there exist a multitude of standards, each potentially representing tens of thousands of unique codes.

In fact, it is the expansiveness of the resulting feature space that leads us to the next aspect of healthcare data that has been shown to impact modeling: sparsity. Although we may be able to capture, code, and quantify an increasingly large feature set, only a small subset of features are often recorded or relevant for a particular individual. Such a point can be best illustrated with the understanding that it is highly improbable a patient will record more than a fraction of the over 100,000 diagnoses and procedures that can now be discretely represented through the ICD-10 standard~\cite{cdc_icd10}.
 
Finally, taking the considerations of dimensionality and sparsity to the next logical step, we find the concept of imbalance. Building on the notion that for any single instance, the data captured likely represents only a sparse set of values with respect to the possible set of data elements, we must acknowledge that these same features are often used as the response variable for many machine learning applications. Whether the prediction of a future diagnosis, or the readmission probability of patient, effective utilization of data in which only a minority of individuals present the attribute of interest often requires additional processing, or specialized models. 
 
\subsection{Implications to Machine Learning}
Having illustrated a number of processes from which data challenges arise within the healthcare domain, we again address how these challenges, at the modeling stage, impact machine learning algorithms.
 
\subsubsection{Heterogeneity}

From a technical perspective, one of the primary considerations in the application of machine learning methodologies to the increasingly heterogeneous healthcare data space comes with the acknowledgment that the data captured across each source may span a range of data types. Such data can represent categorical/discrete (laboratory SRI values), ordinal (pain scales), or continuous (medication dosages) values. However, as noted by Lewis et al., ``traditional statistical methods that assume Gaussian distributions, or engineering methods that assume vector or matrix input do not obviously generalize to datasets comprised of variable-length strings, vectors of real numbers, trees and networks''~\cite{lewis2006support}.

Further, the heterogeneity of the data sources themselves present a challenge for learning algorithms. As the extent of available healthcare data continues to increase, we must also consider the implications of integrating disparate sources. There, of course, exist the practical concerns including record matching and differences in terminology and standards between systems, where a patient may include their middle name on one form but not another or systems may record height as meters or feet. However, the modeling of heterogeneous integrated data sources presents a number of unique concerns, including the ability to reconcile ``dirty" data such as incompatible test results, changes in coded data, and the need to ensure trust between systems that share sensitive data~\cite{diamond2009collecting}. 

Finally, at its core, the siloed data sources present a deeper systemic issue. The lack of a unique identifier to track an individual throughout the various components of the healthcare system often present an incomplete view of any individual's health data~\cite{hillestad2008identity}. While the missing data itself can present concern, the impact of this fragmentation compounds during the integration of multiple data sources. Incorrectly associating records of one patient from two hospitals, records between a primary and specialist, or multiple instances of the same record has serious implications to the machine learning algorithms used to analyze the data. At a basic level, this provides duplicate data that can bias the underlying distributions. While at a higher-level, this removes a true independence assumption, potentially biasing performance measures by splitting what appears to be unique instances amongst the train and test sets during evaluation.

\subsubsection{Dimensionality, Sparsity and Imbalance}

With the considerable advancement of computing systems and hardware over the past few decades, the ability to store, represent, and manipulate high-dimensional data has become commonplace. However, the appropriate utilization of such data in the machine learning pipeline warrants additional consideration. In particular, the scenario often known as the \textit{curse of dimensionality}, in which the number of features approaches or exceeds the number of instances, presents a considerable obstacle to fundamental machine learning theory. One of the most direct impacts results from the emergence of spurious correlation, where many uncorrelated random variables may have high sample correlations~\cite{fan2014challenges}. While more indirectly such a scenario has been shown to breakdown asymptotic theory, preventing the unique estimation of parameters due to occurrence of singular matrices~\cite{johnstone2009statistical}.

Further, the sparsity of feature values presents an additional concern to the development of generalizable models. The identification of latent interactions between features is often not complete, as combinations between all features is rarely captured. While in theory, such an issue can be alleviated with the collection of a large dataset, as feature spaces become ever-larger, the collection of sufficient data is not often feasible due to logistical and economical considerations. While many newer machine learning techniques have looked to sparsity as a foundation for techniques that address the increasing dimensionality using various greedy algorithms, little theoretical support is currently available for such techniques~\cite{lafferty2006challenges}.  

Finally, as noted prior, the presence of sparsity often leads to the traditional class imbalance problem, where the attribute of interest is possessed by only a subset of instances. The implications of such imbalance on statistical learning-tasks have been well established in a number of prior works~\cite{he2009learning}. However it is worth noting a few examples of how the presence of imbalance can impact the modeling of data. From a logistical point of view, a few noisy instances can degrade the identification of the minority class, restricting it to fewer examples to train with. Whereas from a more theoretical presumptive the use of global performance measures that guide the learning process may optimize parameters and decision boundaries in favor of the majority class. Classification rules that predict the positive class are often highly specialized resulting in low coverage of instances across the dataset and may be discarded in favor of providing more general rules~\cite{lopez2013insight}.

\section{Evaluation}
``All models are wrong, some are useful" \cite{box1979robustness}. The provocative and now-famous quote by George Box eloquently provides a fundamental premise of machine learning. The understanding that models may not truly capture the complexity of a system, but rather provide an effective approximation of its observable attributes. This concept has since been defined more formally by Oreskes et al, stating ``Model results may or may not be valid, depending on the quality and quantity of the input parameters and the accuracy of the auxiliary hypotheses" \cite{oreskes1994verification}. In actuality, what these sentiment capture is the need to construct a model in such a way that it accurately represents the system (validation) and accounts for the technical correctness of the model itself (verification)~\cite{kantardzic2011data}.

\subsection{Manifestation in Healthcare}
Together, validation and verification represent a critical aspect of computational tools such as machine learnings impact on our society.  However, accurate assessment of either measure proves a nontrivial task in its own right. Difficulties associated with validation and verification are grounded by two district ideas which govern both measures. First, that there exists a ground truth, and second that any ground truth provided is correct. While both of these assumptions are difficult to guarantee for any real-world data, they are particularly relevant to the variability of data across the healthcare domain. 

The first concern represents a variant of the partially labeled data problem, formally defined by Szummer \cite{szummer2002learning}. As noted in the \textit{Preprocessing} section, although missing values are commonly treated as negative, such an assumption is dangerous. Diagnoses that are never recorded or identified can present the scenario in which two individuals may appear the same from a record standpoint, but in fact present significantly different clinical outcomes. Further, there is a well-established issue in the ability to record clinical conditions completely within any particular coding languages \cite{gensinger2014analytics}. Thus, in an effort to address this lack of a ground truth for missing diagnoses, many machine learning approaches utilize only positive entities, in essence, formulating a one-class problem.

This in turn leads to the second point of concern, where even in the scenario in which a diagnosis is recorded, such data is not guaranteed to be correct. As diagnosis information is often obtained though a diagnostic test, it is important to remember these tests are subjective. Each test is associated with its own performance range quantified by metrics such as sensitivity and specificity, predictive values, chance-corrected measures of agreement, or likelihood ratios~\cite{glas2003diagnostic}. Without the ability to identify false positives/negatives, or to link to follow-up tests or corrections, these results represent a significant source of error which can propagate through an algorithmic model.

\subsection{Implications to Machine Learning}

The challenges associated with the validation and verification of healthcare data impact far more than application-specific performance, reaching to the intermittent steps of the algorithms underlying the solutions themselves. In particular, the complexities of healthcare data impacts both the internal distance metric utilized, as well as the generalized optimization problem of parameter tuning. 

Although simplistic in definition, the notion of distance represents a fundamental attribute in the execution of machine learning algorithms. The ability to quantify a pair of similar or dissimilar points is a concept utilized in determining decision boundaries, as well as updates for weights. Such a concept has become increasingly important as a result of the heterogeneity exhibited by the growing set of healthcare data discussed prior. While work by Brian Kulis highlights extensions of the metric learning problems to a variety of problems in computer vision, text analysis, and multimedia, these extensions represent nontrivial altercations to the concept of what constitutes \textit{distance} \cite{kulis2013metric}.

From an optimization standpoint, in an effort to more accurately model complex real-world systems, an increasingly complex set of learning models have become available. Although these models have the potential to improve performance, there often exist a number of parameters (regularization strengths, number of weak learners, slack variables) that must be tuned to fit the data being modeled \cite{arcuri2013parameter, hoos2011automated}. While specifics of their implementation can vary greatly, to prevent the need to sweep the entirety of the parameter space, a model's parameters can be estimated through optimization techniques applied to an objective function of the users choosing. As there exist a range of possible optimizations, many stemming from a methodology known as gradient decent, an understanding of the data itself is paramount. Variability in how the specified objective function accounts for factors such as class imbalance or inter-feature correlation can result in markedly different results drawn from the same data~\cite{kelley1999iterative, sra2012optimization, lange2014brief}.

Finally, it is important to note that validation and verification have often constituted major components in the assessment of model output, for both its generalizability and overall correctness. While critical to their real-world utility, these concerns directly impact the inferences drawn from model output, and as such, exist outside the scope of this chapter. An excellent survey addressing such items can be found in the work by Sokolova and Lapalme \cite{sokolova2009systematic}.

\section{Open Problems}

This work has served as a foundation highlighting a broad set of considerations that must be addressed as machine learning works to establish its place in the healthcare industry. However, as clinical practice, administration, and research becomes increasingly intertwined with the statistical methodologies of machine learning, there remains a number of open problems. In the sections to follow, we discuss a subset of the most pressing and active areas of research.

\subsection{Temporal Relations}

As healthcare data has undergone the transition from a paper-based entity to digital records, much of the focus has fallen to the purely technical aspects of storing, processing, and modeling such a complex set of variables. Amongst these considerations, however, we often forget to reflect on whether the data we consume accurately reflects the processes it captures. In particular, we find that an overwhelming majority of works have recorded and modeled the condition of an individual as a set of discrete observations.

Although such an approach allows for data to be easily consumed by traditional machine learning approaches, formalizing the presence of each entity recorded as a feature, such a representation is incomplete. It is clear observations such as a diagnosis or procedure must occur at a single point in time, however, it would be naive to believe that such elements of health occur in isolation. Rather, there exist temporal relations connecting them, representing the variable nature of an individual's health. These relations cannot be described by one feature or a single value, but require longitudinal observations with a series of values over time~\cite{zhao2017learning}.

However, this is a nontrivial task from both a computational and clinical standpoint. Taking the example of a patients diagnosis history, the computational complexity of tracking the progression of multiple concurrent diagnoses can quickly become intractable on even the largest of systems, while from a clinical standpoint, many diagnoses have no direct progression, where others may split amongst a broad set of co-morbid diagnoses.

\subsection{Alternative Representations}

Building on the notion that continued improvement of machine learning approaches to problems in the healthcare domain may require a shift in our data organization, such as the ability to capture temporal relations, there exists a significant effort to employ varying frameworks to represent the complexity of healthcare data. Perhaps the most well explored representation can be found in the application of computational networks. Networks represent an established field of interdisciplinary research and present an effective method to capture the direct relation between two arbitrary features, be that a connection between individuals themselves, links between comorbid diagnoses, or genomic-phenotypic relations~\cite{carter2013genotype,feldman2016insights}. 
Such a representation offers many attractive properties, such as the ability to alleviate sparsity by connecting only those elements associated with an instance, and the ability to represent heterogeneous data. However, the analytic methods applied to networks focus primarily on describing connectivity, with measures such as centrality, degree, and betweenness, rather than the generative or discriminative models constructed to describe the relations between various healthcare features. 

In an effort to capture such relations, tensor representations have emerged. A tensor is a multidimensional array spanning an arbitrary number of dimensions, each representing a single feature or modality. Such a representation is advantageous to many areas of healthcare including the ability to capture a series of observations over time or integrate data across multiple experimental conditions and analyze them simultaneously~\cite{hunyadi2016power}. Although tensors are only emerging in their application to healthcare, mathematical operations known as decompositions have demonstrated their value in discovering latent groups in each modality and identify group-wise interactions~\cite{luo2017tensor}.

\subsection{Integration with Clinicians and Clinical Workflows}

Despite the immense technological advancements in the collection and processing of health data, it is important to remember that no system can succeed on its own. We would be remiss in failing to highlight that the impact of machine learning in healthcare cannot be discussed in isolation. Medical research is itself an evolving field, and an understanding of the biological processes being modeled requires a more technical approach. Truly capturing such phenomena will require close interdisciplinary collaborations with those individuals whose expertise lies in the exploration and discernment of healthcare. 

It is perhaps more appropriate to view the continued development of healthcare informatics as part of the complex system encompassing clinical workflows. In relation to the collection and aggregation of data, it is important to note that any increase in data collection poses tangle logistic concerns to those individuals involved in the care of an individual. For example, while it may be more accurate to assess an individual's condition every minute, such granularity is often not feasible. As a result, there must be an increasing focus on developing innovative ways to utilize data collected as part of existing workflows, rather than demonstrating value with models requiring additional data elements. On the other end of the spectrum, in relation to model output, it is important to remember that regardless of the analytic approach used; a patient's treatment ultimately remains in the hands of their clinician. As such, there must be a concerted effort to provide appropriate context to the results. Designing approaches with the capability to quantify factors such as confidence, highlighting a systems strengths, and its weaknesses, for the individual consuming the information. 

\section{Conclusion and Future Outlook}
Complexity comes in many forms, and beyond the sheer volume of data created, the attributes of healthcare data present a vast set of heterogeneous, high-dimensional, probabilistic, incomplete, uncertain, and noisy attributes. However, in conjunction with machine learning methodologies, the increasing availability of data has the potential to provide novel and actionable insights to the field of healthcare.

With this in mind, it is important to remember that although we may have reached a point computationally in which the manipulation of Big Data and execution of complex modeling techniques are possible, purely possessing such capability is not sufficient to ensure the realization of informatics true potential in healthcare. The ability to address such complexities and ensure the effective consumption of healthcare data into the machine learning pipeline relies on more than analytic capability: it relies on a deep understanding of the biological and clinical mechanisms through which the data has been generated.

From preprocessing, to modeling, to the interpretation of model output, this work has presented a general discussion regarding the fundamental challenges presented by such data with respect to the informatics pipeline. However, awareness of such an impact is only the first step. It is our hope that others will draw on these caveats and look to considerations in the design and implementation of new works, which blend together both technological capability and medical understanding to better serve those individuals in need. 

\bibliographystyle{splncs}
\makeatletter
\renewcommand\@biblabel[1]{#1. }
\makeatother

\bibliography{References}

\end{document}